\documentclass{jetpl}

\usepackage{graphicx}

\twocolumn \lat

\begin{document}
\title{Chemomagnetism, magnetoconcentration effect and
"fishtail" anomaly in chemically-induced granular superconductors}
\rtitle{Chemomagnetism, magnetoconcentration effect and "fishtail"
anomaly} \sodtitle{Chemomagnetism, magnetoconcentration effect and
"fishtail" anomaly}

\author{S.A. Sergeenkov}

\address{Unit\'a di Ricerca INFM, Universi\'ta di L'Aquila,
Localit\'a Monteluco, 67040 Roio Poggio, Italy\\ and Laboratory of
Theoretical Physics, Joint Institute for Nuclear Research, 141980
Dubna, Russia}

\dates{20 December 2002}{*}

\abstract{Within a 2D model of Josephson junction arrays (created by
2D network of twin boundary dislocations with strain fields acting as
an insulating barrier between hole-rich domains in underdoped
crystals), a few novel effects expected to occur in intrinsically
granular material are predicted including: (i) Josephson
chemomagnetism (chemically induced magnetic moment in zero applied
magnetic field) and its influence on a low-field magnetization
(chemically induced paramagnetic Meissner effect), and (ii)
magnetoconcentration effect (creation of oxygen vacancies in applied
magnetic field) and its influence on a high-field magnetization
(chemically induced analog of "fishtail" anomaly). The conditions
under which these effects can be experimentally measured in
non-stoichiometric high-$T_c$ superconductors are discussed.}


\maketitle

{\bf 1. Introduction.} Recent imaging of the granular structure in
underdoped $Bi_2Sr_2CaCu_2O_{8+\delta}$ crystals~\cite{1}, revealed
an apparent segregation of its electronic structure into
superconducting domains (of the order of a few nanometers) located in
an electronically distinct background. In particular, it was found
that at low levels of hole doping ($\delta
<0.2$), the holes become concentrated at certain hole-rich domains.
(In this regard, it is interesting to mention a somewhat similar
phenomenon of "chemical localization" that takes place in materials,
composed of atoms of only metallic elements, exhibiting
metal-insulator transitions~\cite{2}.) Tunneling between such domains
leads to intrinsic granular superconductivity (GS) in high-$T_c$
superconductors (HTS). Probably one of the first examples of GS was
observed in $YBa_2Cu_3O_{7-\delta }$ single crystals in the form of
the so-called "fishtail" anomaly of magnetization~\cite{3}. The
granular behavior has been related to the 2D clusters of oxygen
defects forming twin boundaries (TBs) or dislocation walls within
$CuO$ plane that restrict supercurrent flow and allow excess flux to
enter the crystal. Indeed, there are serious arguments to consider
the TB in HTS as insulating regions of the Josephson SIS-type
structure. An average distance between boundaries is essentially less
than the grain size. In particular, the networks of localized grain
boundary dislocations with the spacing ranged from $10 nm$ to $100
nm$ have been observed~\cite{3} which produce effectively continuous
normal or insulating barriers at the grain boundaries. It was also
verified that the processes of the oxygen ordering in HTS leads to
the continuous change of the lattice period along TB with the change
of the oxygen content. Besides, a destruction of bulk
superconductivity in these non-stoichiometric materials with
increasing the oxygen deficiency parameter $\delta $ was found to
follow a classical percolation theory~\cite{4}.

In addition to their importance for understanding the underlying
microscopic mechanisms governing HTS materials, the above experiments
can provide rather versatile tools for designing
chemically-controlled atomic scale Josephson junctions (JJs) and
their arrays (JJAs) with pre-selected properties needed for
manufacturing the modern quantum devices~\cite{5,6}. Moreover, as we
shall see below, GS based phenomena can shed some light on the origin
and evolution of the so-called paramagnetic Meissner effect (PME)
which manifests itself both in high-$T_c$ and conventional
superconductors~\cite{7,8} and is usually associated with the
presence of $\pi$-junctions and/or unconventional ($d$-wave) pairing
symmetry. Since recently, much attention has been paid to both
experimental and theoretical study of PME related effects using
specially designed $SFS$-type junctions~\cite{9,10}.

In this Letter, within a 2D model of JJAs (created by a regular 2D
network of TB dislocations), we discuss a possibility of a few novel
interesting effects which are expected to occur in intrinsically
granular non-stoichiometric material. In particular, we shall
consider (i) Josephson chemomagnetism (chemically induced magnetic
moment in zero applied magnetic field) and its influence on a
low-field magnetization (chemically induced PME), and (ii)
magnetoconcentration effect (creation of extra oxygen vacancies in
applied magnetic field) and its influence on a high-field
magnetization (chemically induced analog of "fishtail" anomaly).

{\bf 2. The scenario.} As is well-known, the presence of a
homogeneous chemical potential $\mu$ through a single JJ leads to the
AC Josephson effect with time dependent phase difference $\partial
\phi /\partial t=\mu /\hbar$. In this paper, we will consider some
effects in dislocation induced JJ caused by a local variation of
excess hole concentration $c({\bf x})$ under the chemical pressure
(described by inhomogeneous chemical potential $\mu ({\bf x})$)
equivalent to presence of the strain field of 2D dislocation array
$\epsilon ({\bf x})$ forming this Josephson contact.

To understand how GS manifests itself in non-stoichiometric crystals,
let us invoke an analogy with the previously discussed dislocation
models of grain-boundary Josephson junctions (GBJJs) (see, e.g.,
\cite{11,12} and further references therein). Recall that under
plastic deformation, grain boundaries (GBs) (which are the natural
sources of weak links in HTS), move rather rapidly via the movement
of the grain boundary dislocations (GBDs) comprising these GBs. Using
the above evidence, in the previous paper~\cite{12} we studied
numerous piezomagnetic effects in granular superconductors under
mechanical loading. At the same time, observed~\cite{1,3,13,14,15} in
HTS single crystals regular 2D dislocation networks of oxygen
depleted regions (generated by the dissociation of $<110>$ twinning
dislocations) with the size $d_0$ of a few Burgers vectors, forming a
triangular lattice with a spacing $d\ge d_0$ ranging from $10nm$ to
$100nm$, can provide quite a realistic possibility for existence of
2D Josephson network within $CuO$ plane. Recall furthermore that in a
$d$-wave orthorhombic $YBCO$ crystal TBs are represented by
tetragonal regions (in which all dislocations are equally spaced by
$d_0$ and have the same Burgers vector ${\bf a}$ parallel to $y$-axis
within $CuO$ plane) which produce screened strain fields~\cite{14}
$\epsilon ({\bf x})=\epsilon _0e^{-{\mid{{\bf x}}\mid}/d_0}$ with
${\mid{{\bf x}}\mid}=\sqrt{x^2+y^2}$.

Though in $YBa_2Cu_3O_{7-\delta }$ the ordinary oxygen diffusion
$D=D_0e^{-U_d/k_BT}$ is extremely slow even near $T_c$ (due to a
rather high value of the activation energy $U_d$ in these materials,
typically $U_d\simeq 1eV$), in underdoped crystals (with
oxygen-induced dislocations) there is a real possibility to
facilitate oxygen transport via the so-called osmotic (pumping)
mechanism~\cite{16,17} which relates a local value of the chemical
potential (chemical pressure) $\mu ({\bf x})=\mu (0)+\nabla \mu \cdot
{\bf x}$ with a local concentration of point defects as follows
$c({\bf x})=e^{-\mu ({\bf x})/k_BT}$. Indeed, when in such a crystal
there exists a nonequilibrium concentration of vacancies, dislocation
is moved for atomic distance $a$ by adding excess vacancies to the
extraplane edge. The produced work is simply equal to the chemical
potential of added vacancies. What is important, this mechanism
allows us to explicitly incorporate the oxygen deficiency parameter
$\delta $ into our model by relating it to the excess oxygen
concentration of vacancies $c_v\equiv c(0)$ as follows
$\delta=1-c_v$. As a result, the chemical potential of the single
vacancy reads $\mu _v\equiv \mu (0)=-k_BT\log (1-\delta )\simeq
k_BT\delta $. Remarkably, the same osmotic mechanism was used by
Gurevich and Pashitskii~\cite{14} to discuss the modification of
oxygen vacancies concentration in the presence of the TB strain
field. In particular, they argue that the change of $\epsilon ({\bf
x})$ under an applied or chemically induced pressure results in a
significant oxygen redistribution producing a highly inhomogeneous
filamentary structure of oxygen-deficient nonsuperconducting regions
along GB~\cite{15} (for underdoped superconductors, the vacancies
tend to concentrate in the regions of compressed material). Hence,
assuming the following connection between the variation of mechanical
and chemical properties of planar defects, namely $\mu ({\bf
x})=K\Omega _0\epsilon ({\bf x})$ (where $\Omega _0$ is an effective
atomic volume of the vacancy and $K$ is the bulk elastic modulus), we
can study the properties of TB induced JJs under intrinsic chemical
pressure $\nabla \mu$ (created by the variation of the oxygen doping
parameter $\delta $). More specifically, a single $SIS$ type junction
(comprising a Josephson network) is formed around TB due to a local
depression of the superconducting order parameter $\Delta ({\bf
x})\propto \epsilon({\bf x})$ over distance $d_0$ producing thus a
weak link with (oxygen deficiency $\delta $ dependent) Josephson
coupling $J(\delta )=\epsilon({\bf x})J_0=J_0(\delta )e^{-{\mid{{\bf
x}}\mid}/d_0}$ where $J_0(\delta )=\epsilon _0J_0=(\mu _v/K\Omega _0
)J_0$ (here $J_0\propto \Delta _0/R_n$ with $R_n$ being a resistance
of the junction). Thus, the considered here model indeed describes
chemically induced GS in underdoped systems (with $\delta \neq 0$)
because, in accordance with the observations, for stoichiometric
situation (when $\delta \simeq 0$), the Josephson coupling $J(\delta
) \simeq 0$ and the system loses its explicitly granular signature.

{\bf 3. The model.} To adequately describe chemomagnetic properties
of an intrinsically granular superconductor, we employ a model of 2D
overdamped Josephson junction array which is based on the well known
tunneling Hamiltonian
\begin{equation}
{\cal H}(t)=\sum_{ij}^NJ_{ij}[1-\cos \phi_{ij}(t)]
\end{equation}
and introduces a short-range (nearest-neighbor) interaction between
$N$ junctions (which are formed around oxygen-rich superconducting
areas with phases $\phi _i(t)$), arranged in a two-dimensional (2D)
lattice with coordinates ${\bf x_i}=(x_i,y_i)$. The areas are
separated by oxygen-poor insulating boundaries (created by TB strain
fields $\epsilon({\bf x}_{ij})$) producing a short-range Josephson
coupling $J_{ij}=J_0(\delta )e^{-{\mid{{\bf x}_{ij}}\mid}/d}$. Thus,
typically for granular superconductors, the Josephson energy of the
array varies exponentially with the distance ${\bf x}_{ij}={\bf
x}_{i}-{\bf x}_{j}$ between neighboring junctions (with $d$ being an
average junction size).

If, in addition to the chemical pressure $\nabla \mu ({\bf
x})=K\Omega _0\nabla \epsilon ({\bf x})$, the network of
superconducting grains is under the influence of an applied
frustrating magnetic field ${\bf B}$, the total phase difference
through the contact reads
\begin{equation}
\phi _{ij}(t)=\phi ^0_{ij}+\frac{\pi w}{\Phi _0} ({\bf x}_{ij}\wedge
{\bf n}_{ij})\cdot {\bf B}+\frac{\nabla \mu \cdot {\bf
x}_{ij}t}{\hbar},
\end{equation}
where $\phi ^0_{ij}$ is the initial phase difference (see below),
${\bf n}_{ij}={\bf X}_{ij}/{\mid{{\bf X}_{ij}}\mid}$ with $ {\bf
X}_{ij}=({\bf x}_{i}+{\bf x}_{j})/2$, and $w=2\lambda _L(T)+l$ with
$\lambda _L$ being the London penetration depth of superconducting
area and $l$ an insulator thickness (which, within the discussed here
scenario, is simply equal to the TB thickness~\cite{17}).

To neglect the influence of the self-field effects in a real
material, the corresponding Josephson penetration length $\lambda
_J=\sqrt{\Phi _0/2\pi \mu _0j_c w}$ must be larger than the junction
size $d$. Here $j_c$ is the critical current density of
superconducting (hole-rich) area. As we shall see below, this
condition is rather well satisfied for HTS single crystals.

\begin{figure}
 \centerline{\includegraphics[width=80mm]{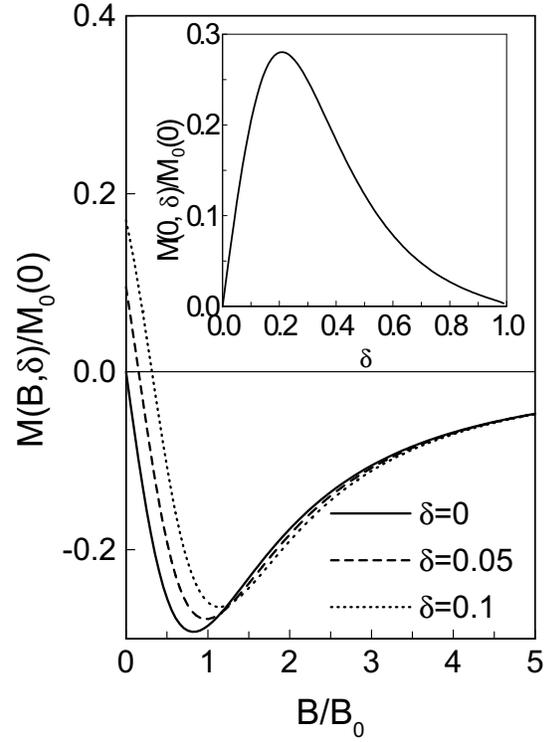}}
 \caption{Fig.\ref{fig:fig1}. The magnetization $M(B,\delta )/M_0(0)$
 as a function of applied magnetic field $B/B_0$, according to
 Eq.(5),
 for different values of oxygen deficiency parameter: $\delta \simeq 0$ (solid line),
 $\delta =0.05$ (dashed line), and $\delta=0.1$ (dotted line).
 Inset: $\delta$ induced magnetization $M(0,\delta )/M_0(0)$ in a
 zero applied magnetic field (chemomagnetism).}
 \label{fig:fig1}
\end{figure}

{\bf 4. Chemomagnetism.} Within our scenario, the sheet magnetization
${\bf M}$ of 2D granular superconductor is defined via the average
Josephson energy of the array
\begin{equation}
<{\cal H}>=\int_0^\tau \frac{dt}{\tau}\int \frac{d^2x}{s} {\cal
H}({\bf x},t)
\end{equation}
as follows
\begin{equation}
 {\bf M}({\bf B},\delta )\equiv -\frac{\partial
<{\cal H}>}{\partial {\bf B}},
\end{equation}
where $s=2\pi d^2$ is properly defined normalization area, $\tau$ is
a characteristic Josephson time, and we made a usual substitution
$\frac{1}{N}\sum_{ij}A_{ij}(t) \to \frac{1}{s}\int d^2x A({\bf x},t)$
valid in the long-wavelength approximation~\cite{18}.

To capture the very essence of the superconducting analog of the
chemomagnetic effect, in what follows we assume for simplicity that a
{\it stoichiometric sample} (with $\delta \simeq 0$) does not possess
any spontaneous magnetization at zero magnetic field (that is
$M(0,0)=0$) and that its Meissner response to a small applied field
$B$ is purely diamagnetic (that is $M(B,0)\simeq -B$). According to
Eq.(4), this condition implies $\phi _{ij}^0=2\pi m$ for the initial
phase difference with $m=0,\pm 1, \pm 2,..$.

Taking the applied magnetic field along the $c$-axis (and normal to
the $CuO$ plane), that is ${\bf B}=(0,0,B)$, we obtain finally
\begin{equation}
M(B,\delta )=-M_0(\delta )\frac{b-b_{\mu }}{(1+b^2)(1+(b-b_{\mu
})^2)}
\end{equation}
for the chemically-induced sheet magnetization of the 2D Josephson
network.

Here $M_0(\delta )=J_0(\delta )/B_0$ with $J_0(\delta )$ defined
earlier (in what follows, $M_0(0)$ is $M_0(\delta \simeq 0)$),
$b=B/B_0$, and $b_{\mu }=B_{\mu }/B_0\simeq (k_BT\tau /\hbar )\delta
$ where $B_{\mu }(\delta )=(\mu _v\tau /\hbar )B_0$ is the
chemically-induced contribution (which disappears in optimally doped
systems with $\delta \simeq 0$), and $B_0=\Phi _0/wd$ is a
characteristic Josephson field.

Fig.~\ref{fig:fig1} shows changes of the initial (stoichiometric)
diamagnetic magnetization $M/M_0$ (solid line) with oxygen deficiency
$\delta$. As is seen, even relatively small values of $\delta$
parameter render a low field Meissner phase strongly paramagnetic
(dotted and dashed lines). The inset of Fig.~\ref{fig:fig1} presents
a true {\it chemomagnetic} effect with concentration (deficiency)
induced magnetization $M(0,\delta )$ in zero magnetic field.
According to Eq.(5), the initially diamagnetic Meissner effect turns
paramagnetic as soon as the chemomagnetic contribution $B_{\mu
}(\delta )$ exceeds an applied magnetic field $B$. To see whether
this can actually happen in a real material, let us estimate a
magnitude of the chemomagnetic field $B_{\mu }$.
Typically~\cite{3,14}, for HTS single crystals $\lambda _L(0)\approx
150nm$ and $d\simeq 10nm$, leading to $B_0\simeq 0.5T$. Using $\tau
\simeq \hbar /\mu _v$ and $j_c=10^{10}A/m^2$ as a pertinent
characteristic time and the typical value of the critical current
density, respectively, we arrive at the following estimate of the
chemomagnetic field $B_{\mu }(\delta )\simeq 0.5B_0$ for $\delta
=0.05$. Thus, the predicted chemically induced PME should be
observable for applied magnetic fields $B\simeq 0.5B_0\simeq 0.25T$
(which are actually much higher than the fields needed to observe the
previously discussed~\cite{12} piezomagnetism and stress induced PME
in high-$T_c$ ceramics). Notice that for the above set of parameters,
the Josephson length $\lambda _J\simeq 1\mu m$, which means that the
assumed in this paper small-junction approximation (with $d\ll
\lambda _J$) is valid and the so-called "self-field" effects can be
safely neglected.

\begin{figure}
 \centerline{\includegraphics[width=75mm]{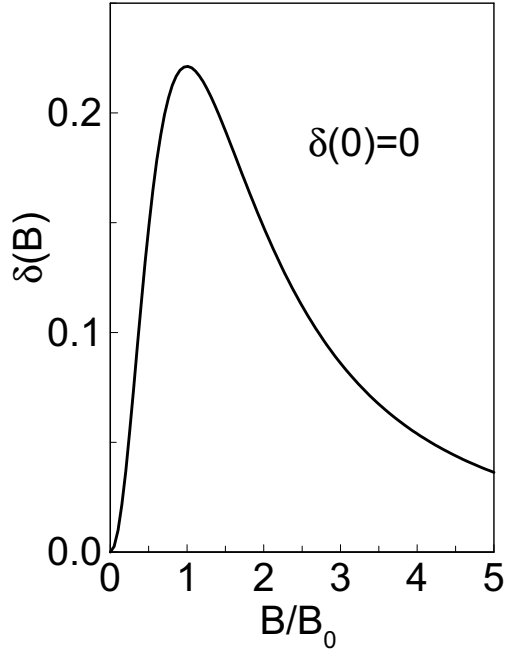}}
 \caption{Fig.\ref{fig:fig2}. Magnetic field dependence of the oxygen
 deficiency parameter $\delta (B)$ (magnetoconcentration effect).}
 \label{fig:fig2}
\end{figure}

{\bf 5. Magnetoconcentration effect.} So far, we neglected a possible
field dependence of the chemical potential $\mu _v$ of oxygen
vacancies. However, in high enough applied magnetic fields $B$, the
field-induced change of the chemical potential $\Delta \mu
_v(B)\equiv \mu _v(B)-\mu _v(0)$ becomes tangible and should be taken
into account. As is well-known~\cite{19,20}, in a superconducting
state $\Delta \mu _v(B)=-M(B)B/n$, where $M(B)$ is the corresponding
magnetization, and $n$ is the relevant carriers number density. At
the same time, within our scenario, the chemical potential of a
single oxygen vacancy $\mu _v$ depends on the concentration of oxygen
vacancies (through deficiency parameter $\delta $). As a result, two
different effects are possible related respectively to magnetic field
dependence of $\mu _v(B)$ and to its dependence on magnetization $\mu
_v(M)$. The former is nothing else but a superconducting analog of
the so-called {\it magnetoconcentration} effect (which was predicted
and observed in inhomogeneously doped semiconductors~\cite{21}) with
field-induced creation of oxygen vacancies $c_v(B)=c_v(0)\exp(-\Delta
\mu _v(B)/k_BT)$, while the latter (as we shall see in the next
Section) results in a "fishtail"-like behavior of the magnetization.
Let us start with the magnetoconcentration effect.
Figure~\ref{fig:fig2} depicts the predicted field-induced creation of
oxygen vacancies $\delta (B)=1-c_v(B)$ using the above-obtained
magnetization $M(B,\delta )$ (see Fig.~\ref{fig:fig1} and Eq.(5)). We
also assumed, for simplicity, a complete stoichiometry of the system
in a zero magnetic field (with $\delta (0)=1-c_v(0)=0$). Notice that
$\delta (B)$ exhibits a maximum at $\delta _c\simeq 0.23$ for applied
fields $B=B_0$ (in agreement with the classical percolative behavior
observed in non-stoichiometric $YBa_2Cu_3O_{7-\delta }$
samples~\cite{3,4,15}). Finally, let us show that in underdoped
crystals the above-discussed osmotic mechanism of oxygen transport is
indeed much more effective than a traditional diffusion. Using
typical $YBCO$ parameters~\cite{14}, $\epsilon _0=0.01$, $\Omega
_0=a_0^3$ with $a_0=0.2nm$, and $K=115GPa$, we have $\mu
_v(0)=\epsilon _0K\Omega _0\simeq 1meV$ for
 a zero-field value of the chemical potential in
HTS crystals, which leads to creation of excess vacancies with
concentration $c_v(0)=e^{-\mu _v(0)/k_BT}\simeq 0.75$ (equivalent to
a deficiency value of $\delta (0)\simeq 0.25$) at $T=T_c$, while the
probability of oxygen diffusion in these materials (governed by a
rather high activation energy $U_d\simeq 1eV$) is extremely slow
under the same conditions because $D\propto e^{-U_d/k_BT_c}\ll 1$. On
the other hand, the change of the chemical potential in applied
magnetic field can reach as much as~\cite{20} $\Delta \mu _v(B)\simeq
0.5meV$ for $B=0.5T$, which is quite comparable with the
above-mentioned zero-field value of $\mu _v(0)$.

\begin{figure}
 \centerline{\includegraphics[width=80mm]{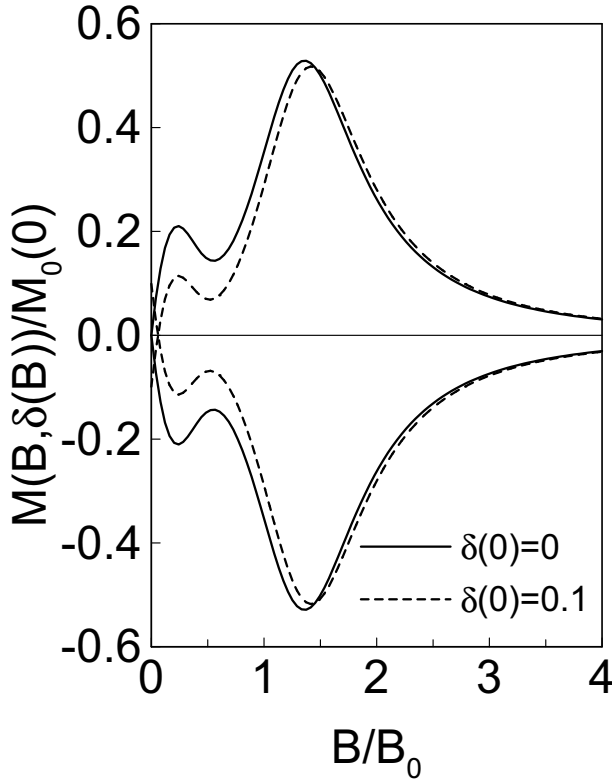}}
 \caption{Fig.\ref{fig:fig3}. A "fishtail"-like behavior of magnetization
 $m_f=M(B,\delta (B))/M_0(0)$ in applied magnetic field $B/B_0$ in the presence
 of magnetoconcentration effect (with field-induced oxygen vacancies
 $\delta (B)$, see Fig.2) for two values of field-free deficiency parameter:
 $\delta (0)\simeq 0$ (solid line), and $\delta
(0)=0.1$ (dashed line).}
 \label{fig:fig3}
\end{figure}

{\bf 6. Analog of "fishtail" anomaly.} Let us turn now to the second
effect related to the magnetization dependence of the chemical
potential $\mu _v(M(B))$. In this case, in view of Eq.(2), the phase
difference will acquire an extra $M(B)$ dependent contribution and as
a result the r.h.s. of Eq.(5) will become a nonlinear functional of
$M(B)$. The numerical solution of this implicit equation for the
resulting magnetization $m_{f}=M(B,\delta (B))/M_0(0)$ is shown in
Fig.~\ref{fig:fig3} for the two values of zero-field deficiency
parameter $\delta (0)$. As is clearly seen, $m_{f}$ exhibits a
field-induced "fishtail"-like behavior typical for underdoped
crystals with intragrain granularity (for symmetry and better visual
effect we also plotted $-m_{f}$ in the same figure). The extra
extremum of the magnetization appears when the applied magnetic field
$B$ matches an intrinsic chemomagnetic field $B_{\mu}(\delta (B))$
(which now also depends on $B$ via the above-discussed
magnetoconcentration effect). Notice that a "fishtail" structure of
$m_{f}$ manifests itself even at zero values of field-free deficiency
parameter $\delta (0)$ (solid line in Fig.~\ref{fig:fig3}) thus
confirming a field-induced nature of intrinsic
granularity~\cite{1,3,13,14,15}. At the same time, even a rather
small deviation from the zero-field stoichiometry (with $\delta
(0)=0.1$) immediately brings about a paramagnetic Meissner effect at
low magnetic fields. Thus, the present model predicts appearance of
two interrelated phenomena, Meissner paramagnetism at low fields and
"fishtail" anomaly at high fields. It would be very interesting to
verify these predictions experimentally in non-stoichiometric
superconductors with pronounced networks of planar defects.

This work was done during my stay in L'Aquila and was funded by the
Italian Institute for the Physics of Matter (INFM). I thank Giacomo
Rotoli and Giovanni Filatrella for hospitality and interesting
discussions on the subject.

\end{document}